# Disintegration of graphene nanoribbons in large electrostatic fields

Haiming Huang[1], Zhibing Li[1,*], H. J. Kreuzer[2] and Weiliang Wang[1,*]

1 State Key Laboratory of Optoelectronic Materials and Technologies, School of Physics and Engineering, Sun Yat-sen University, Guangzhou, 510275, People's Republic of China

2 Department of Physics and Atmospheric Science, Dalhousie University, Halifax, NS B3H 3J5 Canada

[Abstract] The deformation and disintegration of a graphene nanoribbon under external electrostatic fields are investigated by first principle quantum mechanical calculations to establish its stability range. The zigzag edges terminated by various functional groups are considered. By analyzing the phonon spectrum, the critical fracture field for each edge structure is obtained. It is found that different terminal groups on the zigzag graphene nanoribbons lead to different fracture patterns at different fracture fields. The failure mechanism is demonstrated to rely on both the carbon bond alternation feature across the ribbon and terminal group electronegativity.

## 1. INTRODUCTION

Graphene, the thinnest freestanding two-dimensional material found to date[1, 2], has attracted much interest for its unique electronic and magnetic properties[3-5], as well as its exceptional mechanical properties[6-17]. Potential applications of graphene as micro- and nano-electromechanical devices are only starting to emerge[18]. The mechanical properties of different morphological patterns of carbon nanostructures, e.g., one-dimensional graphene nanoribbon (GNR), subjected to external stress have also been studied. It is found that mechanical properties such as Young's modulus and Poisson's ratio of GNR under uniaxial tension can depend strongly on its size and chirality[19]. The study of elastic and plastic deformation of GNRs under uniaxial tension predicts that they attain new functionalities by changing to a range of new structures with interesting electronic and magnetic properties[20]. Furthermore, atomistic simulations showed that the fracture strength decrease only weakly with the width of the GNR flake under tensile loads[21].

---

* Corresponding author. Tel: +86 20 84111107; E-mail address: stslzb@mail.sysu.edu.cn (Zhibing Li)
* Corresponding author. Tel: +86 20 84111107; E-mail address: wangwl2@mail.sysu.edu.cn (Weiliang Wang)

Several studies have predicted that spin-polarized zigzag graphene nanoribbons (ZGNRs)[22-27] under applied electric fields have exceptional electronic and structural properties, e. g., the magnetic properties of ZGNRs can be controlled with external electric fields applied across the zigzag edges and they can become semi-metallic[28]. The alignment and longitudinal polarizability of ZGNRs can also be controlled with external electrostatic fields[29]. Also, the electrocaloric effect of ZGNRs can be enhanced by applying a longitudinal electric field and a transversal magnetic field[30]. In addition, the electric field can function as a switch for uptake/release of adsorbates for storage in graphene[31-33].

We are interested in the stability of ZGNR in strong electrostatic fields, in particular to establish the threshold electrostatic field beyond which GNR will disintegrate.

We will use density functional theory to investigate the structural deformation and the phonon spectrum of ZGNRs with different edge terminations and various widths under electrostatic fields. The critical fracture field is determined by the field strength at which the soft phonon mode starts to appear. The characteristic features of field-induced fragmentation process will be demonstrated by the bond vibration amplitudes. In Sec. 2, we describe in detail the physical model and the computational methodology. The results and their interpretations are presented in Sec. 3, and conclusions are drawn in Sec. 4.

## 2. MODEL AND COMPUTATIONAL METHOD

We first calculate the structural deformation of ZGNRs under uniform electrostatic fields across the ribbons [see Figure 1(f)]. The width of ZGNR is labeled by the number of intact zigzag chains ($N_z$)[34]. The ZGNRs studied here have a width of four or eight zigzag chains (unless specifically mentioned), labeled as 4- or 8-ZGNRs, respectively. We consider five kinds of chemical edge terminations: single hydrogen (CH), double hydrogen (CH2), single fluorine (CF), ketone (CO), and ether (C2O). Their structures are shown in Figure 1. For each value of electric field strength, the ZGNR is relaxed to guarantee equilibrium.

The calculations are based on spin polarized density functional theory (DFT) with generalized gradient approximation (GGA) for the exchange-correlation potential in the form of Perdew-Burke-Ernzerhof (PBE)[35]. We have used effective core potentials with double-numerical plus polarization (DNP) basis sets. All the calculations are performed using the DMOL3 package[36, 37], and periodic boundary conditions are applied to the one-dimensional structure of ZGNR. We consider the smallest supercell along the ribbon direction (i.e., x direction). In order to eliminate interactions between graphene ribbons, 12 Å of vacuum separation along the direction normal to the ZGNR plane was used. The vacuum separation between adjacent in-plane ZGNRs is wider than 12 Å. The atoms are relaxed without any symmetry constraints. The convergence tolerance in the energy is $10^{-5}$ Ha, and the maximum allowed force and displacement are 0.002 Ha/Å and 0.005 Å, respectively. To calculate the structural deformation under applied uniform field in DMOL3, the static potentials arising from externally applied electric field are introduced by adding a potential term to the Hamiltonian. The added potential is a periodic triangular electric potential which is linear along *y* direction (the direction of the applied field) with a potential jump in the middle of the vacuum gap of each pair of neighboring graphene sheets. The validity of this method has been verified by Delley[38].

The phonon spectrum is calculated after complete structural relaxation for each electrostatic field. The phonon spectrum is obtained by computing a complete second derivative Hessian matrix as implemented in the DMOL3 module. The elements of the Hessian are computed by displacing each atom in the system and computing a gradient vector, this builds a complete second derivative matrix. Since the direction of the externally applied electrostatic field and the direction normal to the graphene plane (i.e. y and z direction) have no periodicity, and the supercell size in the x direction is the same as the unit cell, only the Gamma point of the phonon spectrum need to be calculated.

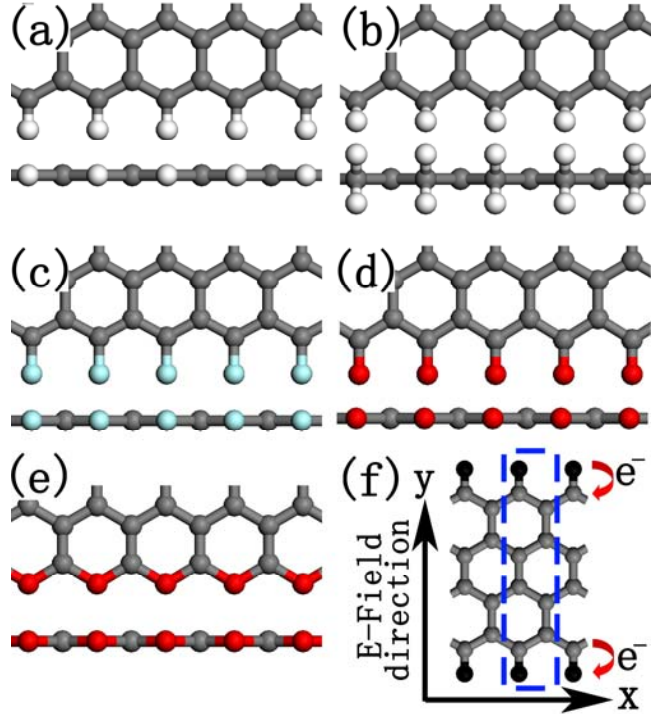

**Figure 1.** Zigzag edge of graphene terminated by (a) single hydrogen (CH), (b) double hydrogen (CH2), (c) single fluorine (CF), (d) ketone (CO) and (e) ether (C2O), respectively. Upper panel: top view; lower panel: side view. (f) Schematic illustration of charge polarization in a zigzag graphene nanoribbon under a uniform applied field across the ribbon. The supercell size of the nanoribbon is indicated by blue dashed rectangle, the height of the rectangle (y direction) is unphysical.

## 3. RESULTS AND DISCUSSION

Before presenting the numerical results, we discuss the physics of the structural deformation of ZGNRs induced by electrostatic fields which are parallel to the graphene planes and normal to the zigzag edges. The functional groups on both edges of each ZGNR are the same. However, when an electrostatic field is applied across the ZGNR, the two edges of the ZGNR are no longer symmetrical due to polarization [Figure 1(f)]. The electrons will move along the direction opposite to that of the applied electric field, making the functional groups on one edge (labeled as front edge) negatively charged while those on the opposite side (labeled as back edge) positively charged. The absolute value of charge on negatively charged functional groups will be larger than that on positively charged functional groups for two reasons: (1) for the functional groups with strong electron-accepting capability (i.e., CF, CO and C2O), the functional groups are negatively charged before the electric field is applied, therefore the absolute value of charge on functional groups on the front edge is larger than that on the back edge which becomes positively charged with strong applied field; (2) for functional groups with weak electronegativity (i.e., CH and CH2), because of the difference in atomic size and in electronic configurations between C atom and the functional groups, the electrons of the C atoms are easier to move which leads to larger absolute value of charge on functional groups on the front edge. We can thus expect that the bond lengths between C atoms and functional groups on both sides will increase with the increasing applied electric field for strong applied fields and the bond lengths between the C atoms and the front edge functional groups will increase faster. This physical picture is confirmed in our calculations, as shown in Figure 2 where the bond lengths and charge of CH and CO 4(8)-ZGNR versus the applied electric field are plotted. For CH 4- and 8-ZGNR [see Figure 2(a), (c)], we can see that the bond lengths between C and H atoms are elongated monotonically; C1–H1 bond length increases obviously faster than C2–H2 bond length under the same applied field, where C1 (C2) and H1 (H2) represent the carbon and hydrogen atoms in the front (back) edge of ZGNR. The bond lengths and charge variation for ZGNR terminated by CH2 show a similar behavior. For CO 4- and 8-ZGNR [see Figure 2(e), (g)], C1–O1 bond lengths are also larger than C2–O2 bond lengths for the same applied field, where C1 (C2) and O1 (O2) represent the carbon and oxygen atoms in the front (back) edge of ZGNR. The minimal bond length is found at the electrostatic field of 1.81(1.43) V/Å for C2-O2 bond of CO 4(8)-ZGNR. This is because the O2 atom is negatively charged at the weak applied fields and becomes positively charged when the field is strong to cause enough charge transfer between C2 and O2 [see Figure 2(f), (h)]. The CF and C2O ZGNR show the same trend as those of CO ZGNR.

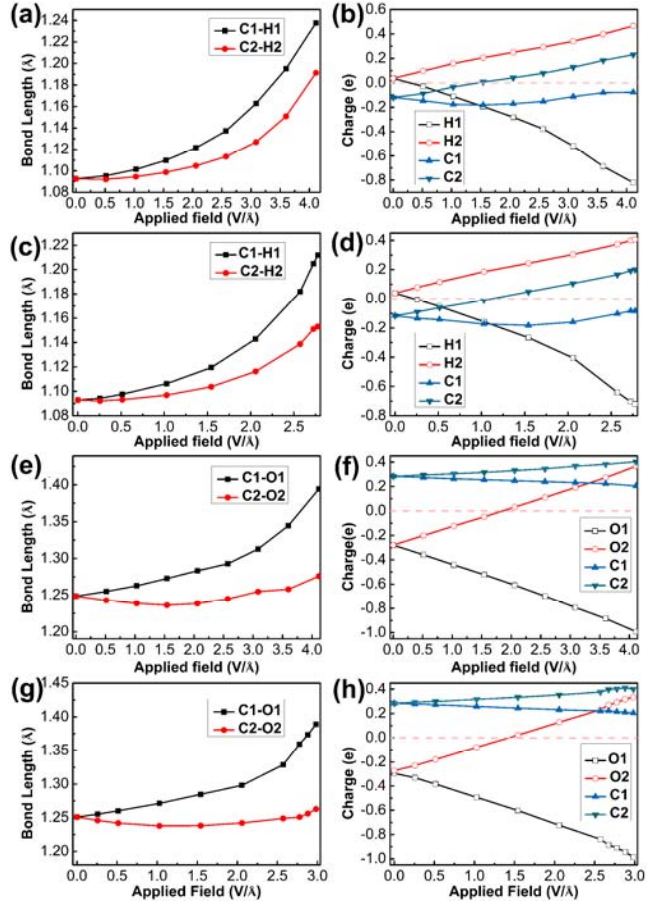

**Figure 2.** The variation of (a) bond lengths and (b) Mulliken charges for CH 4-ZGNR as a function of the applied field, where C1 (C2) and H1 (H2) represent the carbon and hydrogen atoms in the front (back) edge of ZGNR. (c)-(d), (e)-(f) and (g)-(h) are the same as (a)-(b) but for CH 8-, CO 4- and CO 8-ZGNR, respectively.

In order to further investigate the fragmentation and disintegration process of ZGNR, we next explore the phonon spectrum. Figure 3 (a)-(e) show the phonon spectrum of 8-ZGNRs only with that of 4-ZGNRs following same trend. As we are interested in the effect of external electrostatic fields, the optical modes with vibration component in the direction of the external electrostatic field (i.e., y direction) are presented. The monotonic decrease in normal frequency of 8-ZGNR as a function of the applied field indicates the softening of phonons caused by the electrostatic field. It is found that their normal frequencies initially exhibit a slow decrease followed by a sharp decrease as the applied field increases. The ZGNR will be fragmented by the applied field as the normal frequency vanishes. To find the critical fracture electric field, we fit the energy variation of the lowest optical mode with an exponential function $A \cdot \exp(-E_a / B) + C$, where $E_a$ is the applied field and $A$, $B$, $C$ are fitting parameters, as shown in the insets of Figure 3 (a)-(e). The function fits all structures very well. The magnitude of the critical electric fields for ZGNRs with various edge terminations are listed in Table 1. It is found that the critical fracture electric field for wider ZGNR is smaller than that of narrower ZGNR. The reason should be that the interaction between two edges is decreasing with the width. In order to obtain the universal critical fracture electric field for wide ZGNR of which the edge-edge interaction is negligible, we calculated the phonon spectrum for CH ZGNR consisting of 4 to 12 zigzag chains; the results are shown in Figure 3(f). Clearly, the interaction between edges is important for the critical fracture electric field for narrow ZGNR. The critical electric field decreases sharply first and remains nearly constant for width $N_z \geq 8$. Thus the width adopted in the present discussion ($N_z = 8$) is wide enough. Notably, the width dependence of the fracture point here is different from that of the GNR flake under tensile load[21] for which the fracture happen sooner for narrower ribbons. The difference likely comes from the change of mechanical properties caused by charge transfer. Furthermore, from Table 1, it is clear that CH2 ZGNR is the easiest to be fractured with an applied field. This is a consequence of the bond-strength properties of $sp^3$ hybridization. The C2O ZGNR is found to be the most stable structure under the applied field. The fracture electrostatic fields in Table 1 can be routinely achieved by field enhancement effect on the graphene edge. Their magnitudes are comparable to that of microscopic electric field near the apex of field electron emitter as well as in field evaporation and field ion emission[39-43].

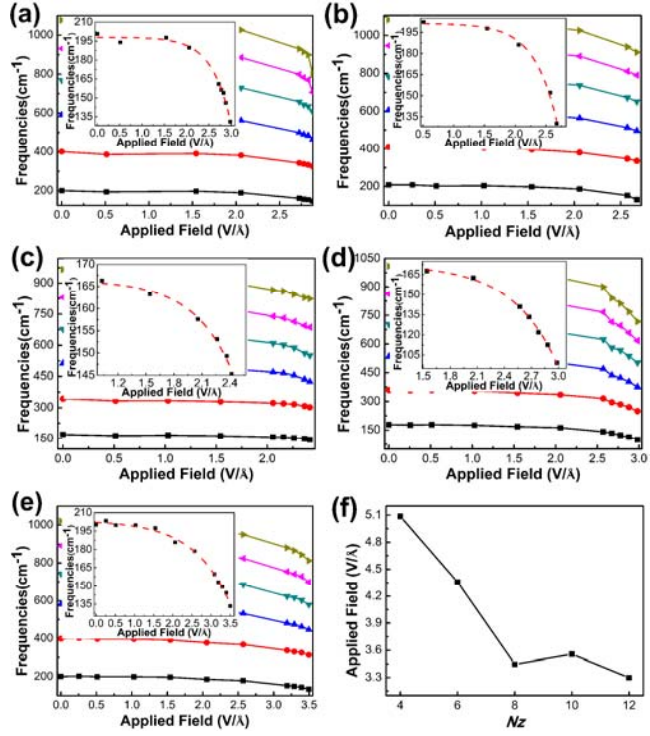

**Figure 3.** (a) The frequency variation of six lowest energy optical modes with vibration component in the direction of the applied electrostatic field for CH 8-ZGNR as a function of the applied field. Inset: a zoom of frequency variation for the lowest energy optical mode. The exponential function fit for variation of the lowest energy optical mode is shown in the inset with dashed line. (b)-(e) The same as (a) but for CH2, CF, CO and C2O 8-ZGNR, respectively. (f) Critical fracture electric field of CH ZGNR versus the width ($N_z$).

**Table 1.** The critical fracture electrostatic fields (V/Å) for 4- and 8-ZGNRs with different edge terminations.

|  | CH | CH2 | CF | CO | C2O |
|---|---|---|---|---|---|
| 4-ZGNR | 5.09 | 4.50 | 4.65 | 5.01 | 5.87 |
| 8-ZGNR | 3.45 | 3.07 | 3.29 | 3.40 | 4.52 |

It remains to discuss how the ZGNRs actually fragment at the critical fracture electric fields. To answer this question, we calculated the vibration amplitude variations (relative to the absence of the applied field) for all chemical bonds in a supercell as a function of the applied field, as shown in left panels of Figure 4. The initial structure is optimized in the absence of the applied field. We focus on the vibration amplitude variations of chemical bonds of the lowest optical mode since its normal frequency decreases to zero firstly (see Figure 3). From left panels of Figure 4, we see that the vibration amplitude of bonds parallel to the direction of the applied field [namely, chemical bonds with odd number in left panels of Figure 4(a)-(d) and even number in left panel of Figure 4(e), labeled as lateral bonds] is stretched more than those in other directions [namely, chemical bonds with even number in left panels of Figure 4(a)-(d) and odd number in left panel of Figure 4(e), labeled as tilted bonds]. For CH2 and C2O 8-ZGNR, the increase of vibration amplitudes of chemical bonds is nearly the same for all lateral bonds [Figure 4 (b), (e)], indicating that CH2 and C2O 8-ZGNRs will be broken up into parts of random size. For CO 8-ZGNR, the change of the C-C lateral bond in the left middle of the ribbon is more significant [Figure 4 (d)], indicating that it will be fractured in the left middle of the ribbon (near the front edge). We also find that the increase of vibration amplitudes for lateral bonds near the front edge is more pronounced than those near the back edge for CH and CF 8-ZGNR [Figure 4 (a), (c)]. Thus the applied field will first break chemical bonds of the front edge. In particular for CF 8-ZGNR, the vibration amplitude of its F-C bond at the front edge increases significantly as the electric field increases, therefore the outermost row of the zigzag edge would be first fragmented.

Close correlation between the above fracture patterns and carbon bond alternation feature for the ZGNR can be found. The significant feature of single-double carbon bond alternation across the ZGNR has been noticed by Kudin[44]. There are two types of single-double carbon bond alternation which rely on edge termination. For the ZGNRs with edge carbons passivated completely, the lateral (tilted) bonds will be double (single) bonds [CH2, CO and C2O ZGNR in right panel of Figure 4(b, d, e), categorized into class I]; otherwise, the double (single) bonds will lie on tilted (lateral) bonds [CH and CF ZGNR in right panel of Figure 4(a, c), categorized into class II]. The failure sites locate at the lateral bonds inside the domain of ZGNR in class I [Figure 4(b, d, e)], while they locate at the lateral bonds on the front boundary of ZGNR in class II [Figure 4 (a, c)]. The different patterns can be explained by two factors. First, lateral bond is stretched more than tilted bond due to the difference of effective moment along electric field. Second, because of the shorter bond length and stronger bond strength thus the larger pre-strain of double lateral bond than single lateral bond, the double lateral bonds in class I show smaller and smoother amplitude variation while the single lateral bonds in class II show more significant increase of amplitude variation on the front edge. Moreover, the electronegativity of the functional groups also plays an important role in the graphene failure. For stronger electronegativity functional group, the bonds near the front edge will have larger electric moment due to the larger field-induced charge transformation. It results in that the fragmentation starts on the outer row of the zigzag edge [CF ZGNR in Figure 4(c)].

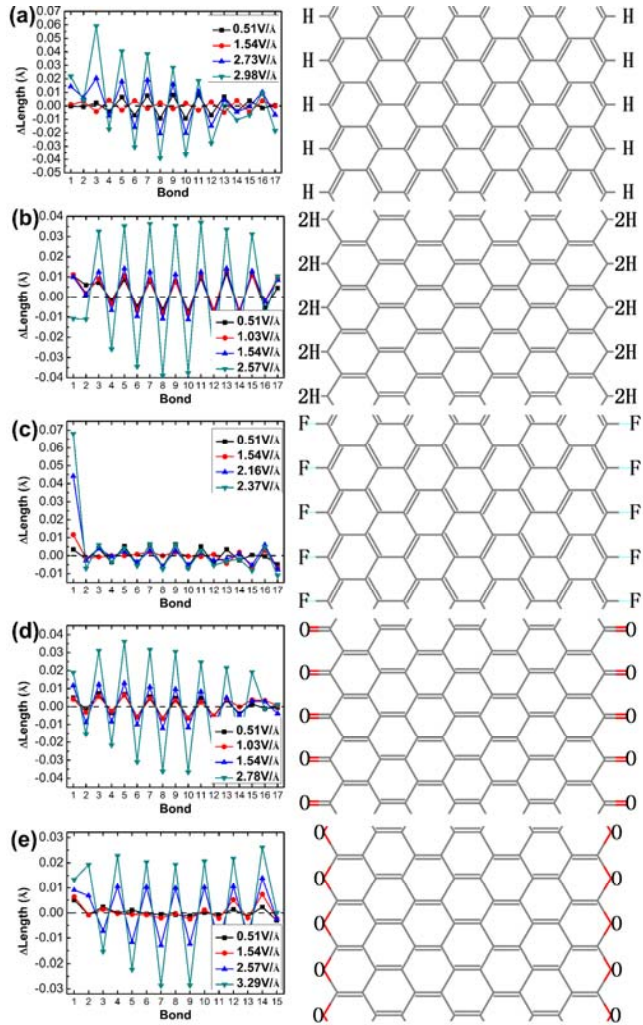

**Figure 4.** (a) Left panel: the vibration amplitude variations (relative to the absence of applied field) of chemical bonds under different electric fields for CH 8-ZGNR. The chemical bonds are labeled by integers along the electric field direction. Right panel: the single-double carbon bond alternation feature across the ribbon of CH 8-ZGNR. (b)-(e) The same as (a) but for CH2, CF, CO and C2O 8-ZGNR, respectively.

## 4. CONCLUSIONS

We investigated the deformation and fragmentation of zigzag graphene nanoribbons with various saturated edges under external uniform electrostatic fields. The charge transfer and the change of bond lengths were calculated. The larger stretching of the terminal bonds in the front edge (higher voltage) than that in the back edge (lower voltage) is explained by the field-induced charge transfer.

We also analyze the phonon spectrum of ZGNRs that is varying with the applied field. It is found that the normal frequencies first decrease smoothly and then exponentially as the electric field increases. The critical electric field for ZGNR fragmentation has been obtained from the vanishing point of the phonon frequency. The stability of ZGNRs saturated with various functional groups under the same applied field follows the hierarchy $C_2O > CH > CO > CF > CH_2$. The critical fracture electric fields for wider ZGNRs are weaker due to the smaller interaction of two polarized edges.

Fragmenting of ZGNRs under the applied field has distinct patterns which depend on edge termination. CH and CF ZGNR can be evaporated from the front edge while CO ZGNR will be torn in the middle; on the contrary, $CH_2$ and $C_2O$ ZGNRs would be broken up into parts randomly. The failure mechanism is demonstrated to rely on both the carbon bond alternation feature across the ribbon and terminal group electronegativity.

Our work will provide useful insights into the way electric fields affect the edge structure; this should be important for the fabrication of nanoelectronic devices. It also suggests a way to identify chirality of graphene edges and shed light on the study of ion emission.

# ACKNOWLEDGMENTS

The project was supported by the National Basic Research Program of China (Grant Nos. 2013CB933601 and 2008AA03A314), the National Natural Science Foundation of China (Grant Nos. 11274393 and 11104358), the Fundamental Research Funds for the Central Universities (No. 13lgpy34), the high-performance grid computing platform of Sun Yat-sen University, the Guangdong Province Key Laboratory of Computational Science, and the Guangdong Province Computational Science Innovative Research Team.

REFERENCES.

[1] Neto AHC, Guinea F, Peres NMR, Novoselov KS, Geim AK. The electronic properties of graphene. Rev Mod Phys 2009;81:109-62.

[2] Geim AK, Novoselov KS. The rise of graphene. Nature Mater 2007;6:183-91.

[3] Karpan VM, Giovannetti G, Khomyakov PA, Talanana M, Starikov AA, Zwierzycki M, et al. Graphite and graphene as perfect spin filters. Phys Rev Lett 2007;99:176602.

[4] Liu X, Pan L, Zhao Q, Lv T, Zhu G, Chen T, et al. UV-assisted photocatalytic synthesis of ZnO-reduced graphene oxide composites with enhanced photocatalytic activity in reduction of Cr(VI). Chem Eng J 2012;183:238-43.

[5] Payne SH, Kreuzer HJ. Collective diffusion in two-dimensional systems: exact analysis based on the kinetic lattice gas model. J Phys-Condens Mat 2009;21:134013.

[6] Zhang J, Zhao J. Mechanical properties of bilayer graphene with twist and grain boundaries. J Appl Phys 2013;113:043514.

[7] Lee C, Wei X, Kysar JW, Hone J. Measurement of the elastic properties and intrinsic strength of monolayer graphene. Science 2008;321:385-8.

[8] Booth TJ, Blake P, Nair RR, Jiang D, Hill EW, Bangert U, et al. Macroscopic graphene membranes and their extraordinary stiffness. Nano Lett 2008;8:2442-6.

[9] Blakslee OL, Proctor DG, Seldin EJ, Spence GB, Weng T. Elastic constants of compression-annealed pyrolytic graphite. J Appl Phys 1970;41:3373-82.

[10] Frank IW, Tanenbaum DM, Van der Zande AM, McEuen PL. Mechanical properties of suspended graphene sheets. J Vac Sci Technol B 2007;25:2558-61.

[11] Gomez-Navarro C, Burghard M, Kern K. Elastic properties of chemically derived single graphene sheets. Nano Lett 2008;8:2045-9.

[12] Tsoukleri G, Parthenios J, Papagelis K, Jalil R, Ferrari AC, Geim AK, et al. Subjecting a Graphene Monolayer to Tension and Compression. Small 2009;5:2397-402.

[13] Liu F, Ming P, Li J. Ab initio calculation of ideal strength and phonon instability of graphene under tension. Phys Rev B 2007;76:064120.

[14] Kudin KN, Scuseria GE, Yakobson BI. C2F, BN, and C nanoshell elasticity from ab initio computations. Phys Rev B 2001;64:235406.

[15] Meo M, Rossi M. Prediction of Young's modulus of single wall carbon nanotubes by molecular-mechanics based finite element modelling. Compos Sci Technol 2006;66:1597-605.

[16] Konstantinova E, Dantas SO, Barone PMVB. Electronic and elastic properties of two-dimensional carbon planes. Phys Rev B 2006;74:035417.

[17] Zhang H, Duan Z, Zhang X, Liu C, Zhang J, Zhao J. Strength and fracture behavior of graphene grain boundaries: effects of temperature, inflection, and symmetry from molecular dynamics. Phys Chem Chem Phys 2013;15:11794-9.

[18] Zhou H, Zhang L, Mao J, Li G, Zhang Y, Wang Y, et al. Template-directed assembly of pentacene molecules on epitaxial graphene on Ru(0001). Nano Res 2013;6:131-7.

[19] Zhao H, Min K, Aluru NR. Size and Chirality Dependent Elastic Properties of Graphene Nanoribbons under Uniaxial Tension. Nano Lett 2009;9:3012-5.

[20] Topsakal M, Ciraci S. Elastic and plastic deformation of graphene, silicene, and boron nitride honeycomb nanoribbons under uniaxial tension: A first-principles density-functional theory study. Phys Rev B 2010;81:024107.

[21] Bu H, Chen Y, Zou M, Yi H, Bi K, Ni Z. Atomistic simulations of mechanical properties of graphene nanoribbons. Phys Lett A 2009;373:3359-62.

[22] Han MY, Oezyilmaz B, Zhang Y, Kim P. Energy band-gap engineering of graphene nanoribbons. Phys Rev Lett 2007;98:206805.

[23] Fujita M, Wakabayashi K, Nakada K, Kusakabe K. Peculiar localized state at zigzag graphite edge. J Phys Soc Jpn 1996;65:1920-3.

[24] Nakada K, Fujita M, Dresselhaus G, Dresselhaus MS. Edge state in graphene ribbons: nanometer size effect and edge shape dependence. Phys Rev B 1996;54:17954-61.

[25] Zeng J, Chen K-Q, He J, Zhang X-J, Sun CQ. Edge Hydrogenation-Induced Spin-Filtering and Rectifying Behaviors in the Graphene Nanoribbon Heterojunctions. J Phys Chem C 2011;115:25072-6.

[26] Miyamoto Y, Nakada K, Fujita M. First-principles study of edge states of H-terminated graphitic ribbons. Phys Rev B 1999;59:9858-61.

[27] Miyamoto Y, Nakada K, Fujita M. First-principles study of edge states of H-terminated graphitic ribbons (vol 59, pg 9858, 1999). Phys Rev B 1999;60:16211-.

[28] Son YW, Cohen ML, Louie SG. Half-metallic graphene nanoribbons. Nature 2006;444:347-9.

[29] Wang Z. Alignment of graphene nanoribbons by an electric field. Carbon 2009;47:3050-3.

[30] Reis MS, Soriano S. Electrocaloric effect on graphenes. Appl Phys Lett 2013;102:112903.

[31] Ao ZM, Hernandez-Nieves AD, Peeters FM, Li S. The electric field as a novel switch for uptake/release of hydrogen for storage in nitrogen doped graphene. Phys Chem Chem Phys 2012;14:1463-7.

[32] Ao ZM, Peeters FM. Electric field: A catalyst for hydrogenation of graphene. Appl Phys Lett 2010;96:253106.

[33] Suarez AM, Radovic LR, Bar-Ziv E, Sofo JO. Gate-Voltage Control of Oxygen Diffusion on Graphene. Phys Rev Lett 2011;106:146802.

[34] Son Y-W, Cohen ML, Louie SG. Energy gaps in graphene nanoribbons. Phys Rev Lett 2006;97:216803.

[35] Perdew JP, Burke K, Ernzerhof M. Generalized gradient approximation made simple. Phys Rev Lett 1996;77:3865-8.

[36] Delley B. An all-electron numerical method for solving the local density functional for polyatomic molecules. J Chem Phys 1990;92:508-17.

[37] Delley B. From molecules to solids with the DMol(3) approach. J Chem Phys 2000;113:7756-64.

[38] Delley B. Vibrations and dissociation of molecules in strong electric fields: N 2, NaCl, H 2O and SF 6. Theochem 1998;434:229-37.

[39] Tsong TT. Atom Probe Field Ion Microscopy. Cambridge: Cambridge University Press; 1990.

[40] Zhang S, Zhang Y, Huang S, Liu H, Wang P, Tian H. First-Principles Study of Field Emission Properties of Graphene-ZnO Nanocomposite. J Phys Chem C 2010;114:19284-8.

[41] Lee SB, Kim S, Ihm J. First-principles dynamic simulations of field emission from carbon nanotubes on gold substrate. Phys Rev B 2007;75:075408.

[42] Ono T, Hirose K. First-principles study on field evaporation for silicon atom on Si(001) surface. J Appl Phys 2004;95:1568-71.

[43] Silaeva EP, Karahka M, Kreuzer HJ. Atom Probe Tomography and field evaporation of insulators and semiconductors: Theoretical issues. Curr Opin Solid St M 2013;17:211-6.

[44] Kudin KN. Zigzag graphene nanoribbons with saturated edges. Acs Nano 2008;2:516-22.